# Experimental Discovery of the First Nonsymmorphic Topological Insulator KHgSb

J.-Z. Ma,[1,†] C.-J. Yi,[1,†] B. Q. Lv,[1,†] Z. J. Wang,[2,†] S.-M. Nie,[1] L. Wang,[1] L.-Y. Kong,[1] Y.-B. Huang,[3] P. Richard,[1,4] H.-M. Weng,[1,4] B. A. Bernevig,[2] Y.-G. Shi,[1,*] T. Qian,[1,4,*] and H. Ding[1,4,*]

[1] *Beijing National Laboratory for Condensed Matter Physics and Institute of Physics, Chinese Academy of Sciences, Beijing 100190, China*

[2] *Department of Physics, Princeton University, Princeton, New Jersey 08544, USA*

[3] *Shanghai Synchrotron Radiation Facility, Shanghai Institute of Applied Physics, Chinese Academy of Sciences, Shanghai 201204, China*

[4] *Collaborative Innovation Center of Quantum Matter, Beijing, China*

[†] These authors contributed to this work equally.

[*] Corresponding authors: dingh@iphy.ac.cn, tqian@iphy.ac.cn, ygshi@iphy.ac.cn

**Topological insulators (TIs) host novel states of quantum matter, distinguished from trivial insulators by the presence of nontrivial conducting boundary states connecting the valence and conduction bulk bands. Up to date, all the TIs discovered experimentally rely on the presence of either time reversal or symmorphic mirror symmetry to protect massless Dirac-like boundary states. Very recently, it has been theoretically proposed that several materials are a new type of TIs protected by nonsymmorphic symmetry, where glide-mirror can protect novel exotic surface fermions with hourglass-shaped dispersion. However, an experimental confirmation of such new nonsymmorphic TI (NSTI) is still missing. Using angle-resolved photoemission spectroscopy, we reveal that such hourglass topology exists on the (010) surface of crystalline KHgSb while the (001) surface has no boundary state, which is fully consistent with first-principles calculations. We thus experimentally demonstrate that KHgSb is a NSTI hosting hourglass fermions. By expanding the classification of topological insulators, this**

**discovery opens a new direction in the research of nonsymmorphic topological properties of materials.**

The discovery of symmetry protected states of matter in time-reversal TIs and topological semimetals [1-9] has attracted tremendous interest in a wide range of areas from condensed matter physics to material science, quantum chemistry and high-energy physics. The first TIs [10] discovered possess time-reversal symmetry with Kramers degeneracy protecting nontrivial conducting states on the boundary of two and three-dimensional samples. Over the past decade, the classification of topological insulating phases has been theoretically extended to many other discrete symmetry classes [11-13]. Amongst these, we mention particle-hole symmetry for topological superconductors, certain magnetic translation symmetry for antiferromagnetic TIs, and crystal symmetries for topological crystalline insulators. The crystalline TIs studied in the past rely on the presence of symmorphic symmetries: the only experimentally discovered class of crystalline TI's rely on the presence of a symmorphic mirror plane. In all the previously known cases time-reversal and/or crystal symmetry generically protects massless boundary states with the usual Dirac-like dispersion [10,14]. Among spatial symmetries, symmorphic symmetries preserve the origin, while nonsymmorphic symmetries unavoidably translate the origin by a rational fraction of the lattice vector. The fractional translation could in principle play an important role in the classification of band topologies without analogue in symmorphic crystals. Very recently, Wang *et al* and Alexandradinata *et al* [15,16] proposed the first theory of nonsymmorphic time-reversal-invariant insulators. In contrast to the surface Dirac fermions ubiquitous in symmorphic TIs, NSTIs exhibit bizarre surface fermions with hourglass-shaped dispersion (Fig. 1a). The authors of [15] also proposed that this new kind of topological insulator can be realized in the series of three materials KHg$X$ ($X$ = As, Sb, Bi).

We start with a brief review of the surface state properties predicted theoretically for KHgSb, which is measured experimentally here. KHgSb is predicted to host an hourglass fermion on the (010) surface (Fig. 1e) within a large global bulk band gap. The crystal structure of KHgSb is characterized by nonsymmorphic space group $D_{6h}^4$

($P6_3/mmc$), which is composed of alternately stacked honeycomb HgSb and hexagonal K layers, with the adjacent HgSb layers rotated by 60° with respect to each other (Fig. 1b). The spatial symmetries of KHgSb include two kinds of nonsymmorphic symmetries: a $C6$ screw rotation with a translation along the rotational axis and a glide reflection $\overline{M}_x$ with a translation parallel to the reflection plane. The glide reflection $\overline{M}_x$ is preserved on the (010) surface, which is crucial to form the unique hourglass fermions on this surface. As illustrated in Fig. 1e, due to time-reversal symmetry, the complex-conjugate $\overline{M}_x$ eigenvalues are paired at time-reversal-invariant momenta $\tilde{\Gamma}$ and $\tilde{Z}$ as $\pm i$ and +1 (or -1) pairs, respectively. The degeneracies at $\tilde{\Gamma}$ and $\tilde{Z}$ have to exchange the glide partners along the glide-symmetric line $\tilde{\Gamma}-\tilde{Z}$, resulting in an hourglass-shaped dispersion along $\tilde{\Gamma}-\tilde{Z}$. The crossing of the hourglass is stabilized by glide symmetry.

To search for hourglass fermions, we performed systematic angle-resolved photoemission spectroscopy (ARPES) measurements to investigate the electronic structure of KHgSb. The chemical composition of KHgSb single crystals is confirmed by core level photoemission measurements. As shown in Fig. 1g, the characteristic peaks of K, Hg, and Sb elements are clearly observed. Our X-ray diffraction (XRD) measurements on the (001) surface of KHgSb show that the lattice constant *c* corresponds to twice the distance between the adjacent HgSb layers (Fig. 1f), confirming the *A-B-A-B*-type double-layered lattice. The low-energy electron diffraction (LEED) pattern on the (001) surface shows a hexagonal structure consistent with the six-fold symmetry of the crystal structure (inset of Fig. 1f).

We first summarize the ARPES results recorded on the (001) surface of KHgSb in Fig. 2. On this surface, the chemical potential lies in the band gap between the valence and conduction bands (Figs. 2a-c), demonstrating the insulating nature of the bulk states in this material. The measured band dispersions along the high-symmetry lines $\overline{\Gamma}-\overline{M}-\overline{K}-\overline{\Gamma}$ are in excellent agreement with the calculated valence band structure (Figs. 2b,c). The valence bands form an "M"-like top along both $\overline{\Gamma}-\overline{M}$ and $\overline{\Gamma}-\overline{K}$. The band top is close to $E_F$, resulting in enhanced spectral intensity in the vicinity of $\overline{\Gamma}$ at $E_F$ (Fig. 2a).

To observe the conduction bands, we deposited potassium atoms onto the (001) surface *in situ*, which dopes electrons into the surface layer and thus shift the local chemical potential upwards. The bottom of the conduction bands is clearly observed around $\bar{\Gamma}$ after the deposition of potassium in Figs. 2d,e. To clarify whether surface states exist or not, we plot the energy distribution curves (EDCs) around $\bar{\Gamma}$ in Figs. 2f,g and fit the EDC at $\bar{\Gamma}$ in Fig. 2h. No sign of surface state is observed within the bulk band gap, in agreement to the predicted theoretical calculations.

We then turn our focus to the ARPES results recorded on the (010) surface where the hourglass-shaped fermion surface state was predicted to exist. While the (010) side-surface is not a natural cleavage surface, we succeeded in obtaining a mirror-like (010 plane several times with many attempts of *in-situ* cleave. Highly dispersive and well-defined bands are clearly observed with good reproducibility. We performed photon energy dependence measurements along $\tilde{\Gamma}-\tilde{X}$ on the (010) surface. These measurements map the electronic structure in the $\Gamma-M-K-\Gamma$ plane of the bulk Brillouin zone (BZ). For the bulk states, the measurements on the (010) surface with varying photon energy should be analogous to the in-plane mapping on the (001) surface at a fixed photon energy. In Fig. 3a, the FS intensity map obtained with varying photon energy on the (010) surface exhibits enhanced spectral intensity near $\Gamma$, which is consistent to our observation on the (001) surface in Fig. 2a. In addition, the FS intensity map in Fig. 3a shows two straight lines at $k_x = 0$ and $2\pi a/\sqrt{3}$ along $\Gamma-M-K-\Gamma$, which is perpendicular to the (010) surface.

To illuminate the origin of the straight-line FSs, we investigate the band dispersions recorded with different photon energies. In Figs. 3b,c, we compare the band dispersions measured along $\Gamma-M$ with the calculated bulk bands. While most of the experimental band dispersions are in good agreement with the bulk band calculations (Fig 3e), we observe an extra feature close to $E_F$ at $k_x = 0$ and $2\pi a/\sqrt{3}$, corresponding to the two straight lines in the FS intensity map. Figure 3d illustrates the evolution of the band dispersions upon sliding along the $k_y$ direction. The valence band top at $\Gamma$ gradually sinks upon moving from $\Gamma$ to $K$, as indicated in Fig. 3d, which is consistent with the bulk band calculations. In contrast, the extra band is observable near $E_F$ in all the cuts in Fig. 3d, though with weaker spectral intensity around $\bar{M}$, due to the photoemission matrix elements. Figure 3f displays the near-$E_F$ band dispersions obtained with different photon energies. The extra band shows a nearly linear dispersion with the crossing point close to $E_F$. The non-dispersive feature of the extra band along $k_y$ (the momentum perpendicular to the surface) indicates its surface origin.

The measurements along $\tilde{\Gamma}-\tilde{X}$ clearly indicate the existence of surface states on the (010) surface, where the hourglass-shaped fermion surface state was predicted to exist. To capture the hourglass fermion and its connectivity to other bands in different directions of the surface BZ, we systematically investigate the low-energy electronic structure of the surface states in the (010) surface BZ in Fig. 4. Let us first review the overall surface band structure in the (010) surface BZ predicted by the calculations. The calculations show two Dirac-like bands that are split in energy along $\tilde{\Gamma}-\tilde{X}$ ($k_z = 0$) leading to two Kramers points with an energy separation of 50 meV at $\tilde{\Gamma}$ (Fig. 4a). These bands are doubly-degenerate along $\tilde{Z}-\tilde{U}$ ($k_z = \pi$) but open a small gap of 10 meV between the upper and lower branches at $\tilde{Z}$ (Fig. 4b). The double-degeneracy of the bands along $\tilde{Z}-\tilde{U}$ is theoretically guaranteed by an anti-unitary symmetry $Tg_x$ - the product of the time reversal and glide symmetries - which yields $(Tg_x)^2 = -1$ at $k_z = \pi$. The crossing points at $\tilde{\Gamma}$ and at $\tilde{Z}$ are nothing else but the doubly-degenerate endpoints of the theoretically calculated hourglass-shaped dispersion along $\tilde{\Gamma}-\tilde{Z}$ in Figs. 4c,d.

Our ARPES results displayed in Figs. 4e,f clearly exhibit one Dirac-like band along $\tilde{\Gamma}-\tilde{X}$ with the crossing point at 150 meV below $E_F$. In addition, we observe one nearly linearly-dispersive band parallel to the lower branch of the Dirac-like band. The observed band dispersions are consistent with the calculated band structure along $\tilde{\Gamma}-\tilde{X}$ in Fig. 4a. Along $\tilde{Z}-\tilde{U}$ we observe only one Dirac-like band with the crossing point at 50 meV below $E_F$ in Figs. 4g,h. The high velocity of the Dirac-like band leads to a large self-energy of the quasiparticles, which hinders the observation of the small gap at $\tilde{Z}$. The energy gap is difficult to identify even in the calculated spectra, which uses a much smaller self-energy (Fig. 4b). The experimental bands along $\tilde{\Gamma}-\tilde{Z}-\tilde{\Gamma}$ (where the hourglass theoretically resides) in Figs. 4f,j exhibit a cross-like dispersion, which suggests that the "pinch" point of the hourglass is very close to the $\tilde{Z}$ point. This observation is well reproduced by the calculations in Fig. 4c, in which the splitting in energy at $\tilde{\Gamma}$ is much larger than that at $\tilde{Z}$. As described above, the bands have to form two Kramers-pairs at both $\tilde{\Gamma}$ and $\tilde{Z}$ that theoretically need to switch partners along $\tilde{\Gamma}-\tilde{Z}$ due to the glide symmetry $\overline{M}_x$, resulting in the hourglass fermions. The hourglass fermions lie slightly below $E_F$, resulting in the enhance intensity along $\tilde{\Gamma}-\tilde{Z}$ in the FS intensity map displayed in Fig. 4k. We summarize in Fig. 4m the measured surface band dispersions on the (010) surface. We clearly observe that the surface hourglass fermions along $\tilde{\Gamma}-\tilde{Z}$ are connected to the bulk states via Dirac-like bands along $\tilde{\Gamma}-\tilde{X}$ and $\tilde{Z}-\tilde{U}$. This observation is consistent with the theoretical prediction.

By mapping a large set of high-symmetry manifolds in the surface BZ, we provided evidence that a newly proposed type of topological insulator, hosting a novel type of surface state, exists in crystals protected by nonsymmorphic symmetry. Our work shows that KHgSb is the first NSTI. This opens a brand new research line in the field of topological insulators: most symmetry groups in nature are nonsymmorphic and can host interesting topological classes with their own unique surface state responses. Lastly, nonsymmorphic symmetries can protect bulk semimetals hosting exotic types of fermions [17]. All these await experimental discovery.

## Acknowledgements

We acknowledge Cheng Fang and Xi Dai for valuable discussions. This work was supported by the Ministry of Science and Technology of China (Nos. 2015CB921300 and 2013CB921700), the National Natural Science Foundation of China (Nos. 11234014, 11274362, 11274367, 11422428, 11474330, 11474340, and 11504117), the Chinese Academy of Sciences (No. XDB07000000). ZJW and BAB were also supported by the ARO MURI W911-NF-12-1-0461, ONR-N00014- 11-1-0635, NSF CAREER DMR-0952428, NSF-MRSEC DMR-1005438, the Schmidt Fund, the Packard Foundation and a Keck grant.

## Author contributions

H.D. and T.Q. conceived the ARPES experiments. J.Z.M., B.Q.L. and T.Q. performed ARPES measurements with the assistance of L.Y.K. and Y.B.H. Z.J.W., S.M.N. and H.M.W. performed *ab initio* calculations. C.J.Y., Y.G.S. and L.W. synthesized the single crystals. J.Z.M., T.Q. and H.D. analyzed the experimental data. J.Z.M., B.Q.L., Z.J.W. and T.Q. plotted the figures. T.Q., H.D., B.A.B., J.Z.M., Z.J.W. and P.R. wrote the manuscript.

## Experimental and Computational Methods

Single crystals of KHgSb have been grown using self-flux methods. The starting materials K (99.95%, Alfa Aesar), Hg (99.9995%, Aladdin) and Sb (antimony shot, 99.999%, Alfa Aesar) were put in a $Al_2O_3$ capsule and sealed in a quartz tube. The operation was in a glove box filled with high-purity argon. We presintered the samples as well by heating at 200 °C for 20 hours in the furnace. After that the samples were sealed in a tantalum capsule and then sealed in a quartz tube, followed by heating at 900 °C for 10 hours and maintaining this temperature for 5 hours. Then the samples were cooled down to 480 °C at a rate of 3 °C/h and hold at this temperature for 5-7 days.

ARPES measurements were performed at the "Dreamline" beamline of the Shanghai Synchrotron Radiation Facility (SSRF) with a Scienta D80 analyser. The energy and angular resolutions were set to 15–30 meV and 0.2°, respectively. The samples for ARPES measurements were mounted in a BIP argon (>99.9999%)-filled glove box and cleaved *in situ* and measured at 30 K in a vacuum better than $5\times10^{-11}$ Torr.

The Vienna *ab-initio* simulation package (VASP) [18] is employed for first-principles calculations. The generalized gradient approximation (GGA) of Perdew-Burke-Ernzerhof type [19] is used for the exchange-correlation potential.

Spin-orbit coupling (SOC) is taken into account self-consistently. The cut-off energy for plane wave expansion is 500 eV and the $k$-point sampling grids are 16×16×8, which have been tested to be dense enough for different structures. The atomic structure and the lattice constants $a = b = 4.78(4)$ Å and $c = 10.22(5)$ Å determined by power diffraction are adapted in our calculations [15]. The KHgSb layers are stacked along the $c$ axis, which is consistent with our experimental observations. The surface Green's function method is used to calculate the surface states, based on the Maximally localized Wannier functions [20] of Hg-6$s$ and Sb-5$p$ orbitals obtained from *ab-initio* calculations.

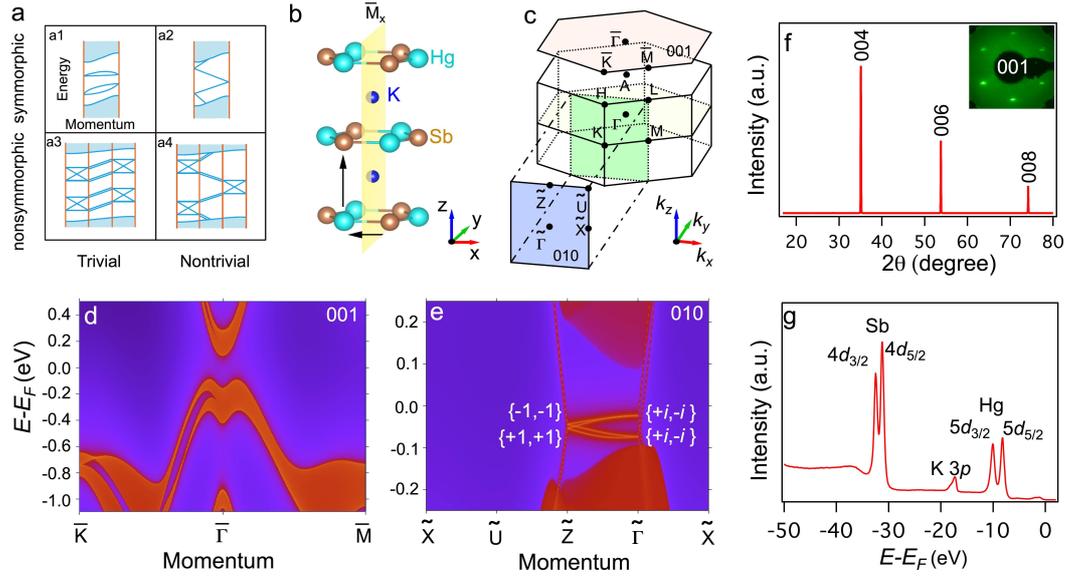

**Figure 1 | Hourglass fermions predicted in KHgSb with nonsymmorphic spatial symmetry**. **a,** Schematic surface state bands for materials with symmorphic and nonsymmorphic spatial symmetries. The shaded region represents bulk states. **a1, a2**: Topologically trivial and nontrivial cases with symmorphic spatial symmetry, respectively. **a3, a4**: Same as **(a1)** and **(a2)** but with nonsymmorphic spatial symmetry showing hourglass-shaped surface band dispersion. **b,** Crystal structure of KHgSb. The yellow region indicates the glide mirror plane $\overline{M}_x$ preserved on the (010) side surface. The arrows indicate the process of the glide reflection $\overline{M}_x$. **c,** 3D bulk BZ of KHgSb as well as its projected (001) and (010) surface BZs. **d, e,** Projections of calculated bulk and surface bands on the (001) and (010) surfaces of KHgSb, respectively. The calculations show hourglass-shaped surface bands along the glide-symmetric line $\tilde{\Gamma}-\tilde{Z}$ on the (010) surface. The complex-conjugate $\overline{M}_x$ eigenvalues are paired as $\pm i$ pairs at $\tilde{\Gamma}$ and +1 (or -1) pairs at $\tilde{Z}$. **f,** XRD measured on the (001) surface of KHgSb crystal confirming the double-layered lattice. Inset: The low-energy electron diffraction pattern shows the hexagonal structure of KHgSb (001) surface. **g,** Core level photoemission spectrum showing characteristic peaks of K 3p, Hg 5d and Sb 4d core levels.

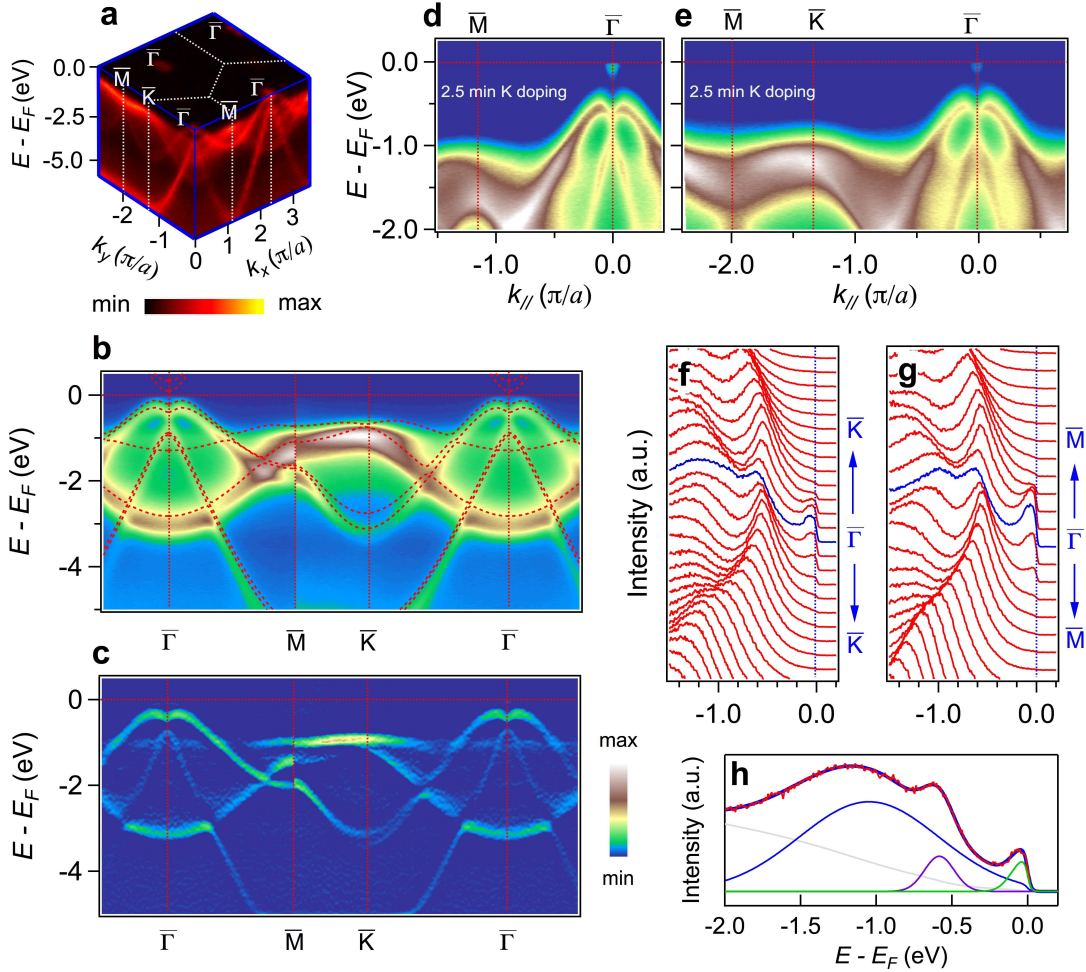

**Figure 2 | Electronic structure on the KHgSb (001) surface.** **a**, 3D intensity plot of ARPES spectra showing the electronic structure of valence bands in the $k_x$-$k_y$ plane. **b,** ARPES intensity plot showing band dispersions along the high symmetry lines $\bar{\Gamma} - \bar{M} - \bar{K} - \bar{\Gamma}$ on the pristine (001) surface. The dashed curves represent the calculated bulk bands at $k_z = 0$. **c,** Corresponding curvature intensity plot of **(b)**. **d, e,** ARPES intensity plots along $\bar{M} - \bar{\Gamma}$ and $\bar{M} - \bar{K} - \bar{\Gamma}$ after *in-situ* K doping, respectively, showing the bottom of the conduction bands. **f, g**, EDCs of **(d)** and **(e)** around $\bar{\Gamma}$, respectively. **h,** Multiple peak fitting of the EDC at $\bar{\Gamma}$, indicating no surface state in the bulk band gap. The black curve represents a Shirley-type background. The blue, magenta, and green curves are the fitted peaks with a Gaussian function multiplied by a Fermi-Dirac function.

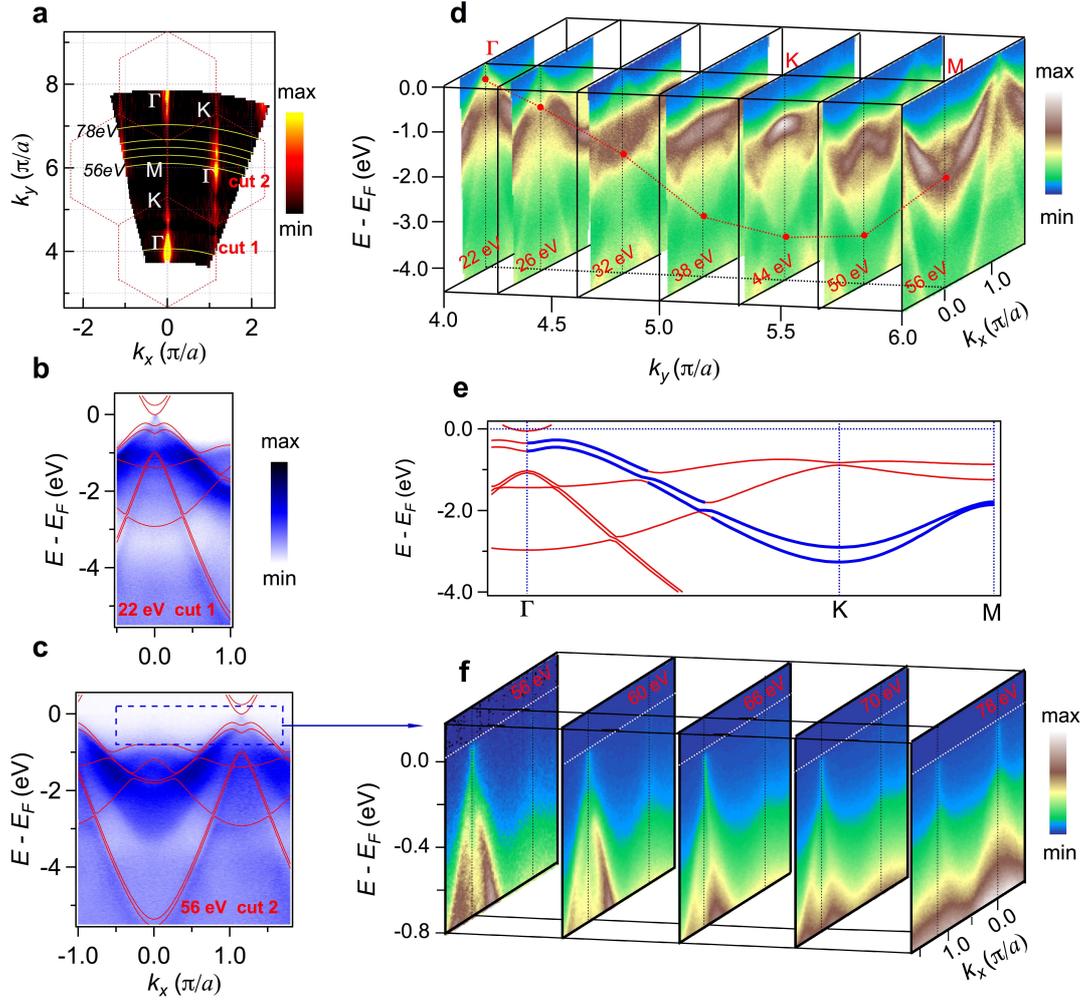

**Figure 3 | Band dispersions along $\tilde{\Gamma} - \tilde{X}$ on the KHgSb (010) surface. a,** ARPES intensity plot at $E_F$ recorded along $\tilde{\Gamma} - \tilde{X}$ by varying the photon energy from 20 to 100 eV on the (010) surface, which maps the $k_x$-$k_y$ plane at $k_z = 0$ of the 3D bulk BZ. The overlaid hexagons indicate the BZ structure in the $k_x$-$k_y$ plane. **b, c,** Band dispersions along cut1 and cut2 indicated in **(a)**, recorded with photon energy $hv = 22$ and 56 eV, respectively. The solid curves represent the calculated bulk bands along $\Gamma - M$. **d,** Band dispersions along $k_x$ at different $k_y$ positions recorded with different photon energies from 22 to 56 eV. The dashed curve indicates the dispersion of one representative bulk band along $\Gamma - K - M$. **e,** Calculated bulk bands along $\Gamma - K - M$. The blue curves indicate the calculated bulk bands that correspond to the experimental band dispersion indicated in **d**. **f,** Near-$E_F$ band dispersions along $k_x$ at different $k_y$ positions recorded with different photon energies from 56 to 78 eV,

showing a Dirac-like surface state band along $\tilde{\Gamma} - \tilde{X}$. The momentum locations are indicated in **(a)**.

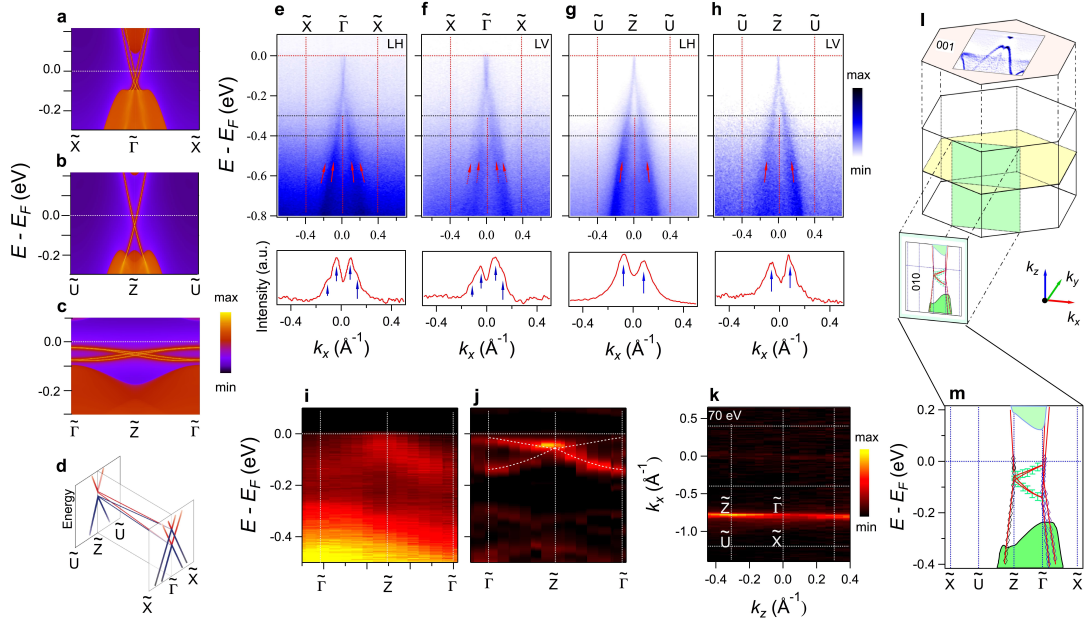

**Figure 4 | Electronic structure of surface states on the KHgSb (010) surface. a-c,** Calculated surface band dispersions along $\tilde{\Gamma}-\tilde{Z}-\tilde{\Gamma}$, $\tilde{X}-\tilde{\Gamma}-\tilde{X}$ and $\tilde{U}-\tilde{Z}-\tilde{U}$, respectively. **d,** 3D sketch of the band structure of hourglass fermions along high symmetry lines. **e-h,** Band dispersions along $\tilde{X}-\tilde{\Gamma}-\tilde{X}$ and $\tilde{U}-\tilde{Z}-\tilde{U}$ recorded with linearly-horizontal (LH) and linearly-vertical (LV) polarized lights, respectively. The momentum distribution curves below each panel in **(e)-(h)** are integrated between the two horizontal dashed lines, indicating that the two Dirac-like bands are split along $\tilde{\Gamma}-\tilde{X}$ and doubly-degenerate along $\tilde{Z}-\tilde{U}$. **i,** Intensity plot of ARPES data along $\tilde{\Gamma}-\tilde{Z}-\tilde{\Gamma}$. **j,** Corresponding curvature intensity plot of **(i)** showing band dispersions along $\tilde{\Gamma}-\tilde{Z}-\tilde{\Gamma}$. The dashed curves indicate the cross-like dispersion. **k,** ARPES intensity plot at $E_F$ recorded at $hv$ = 70 eV on the (010) surface. **l,** Summary of the experimental results showing no surface state on the (001) surface and the existence of hourglass fermions in the bulk band gap on the (010) surface. The intensity plot in the projected (001) surface BZ is the curvature intensity of the ARPES data along $\overline{K}-\overline{\Gamma}-\overline{M}$ in Figs. **1d,e**. The plot in the projected (010) surface BZ is zoomed out from **(m)**. **m,** Surface band structure along $\tilde{X}-\tilde{U}-\tilde{Z}-\tilde{\Gamma}-\tilde{X}$. The symbols represent the extracted bands from the experimental data in **(e)-(j)**. The red curves represent the reconstructed surface band structure based on experiments and calculations. The shaded region represents the bulk states.